\documentclass[pdflatex,sn-mathphys-num,iicol]{sn-jnl}
\usepackage{fix-cm}
\DeclareUnicodeCharacter{2212}{\textminus}

\usepackage{graphicx}%
\usepackage{multirow}%
\usepackage{amsmath,amssymb,amsfonts}%
\usepackage{amsthm}%
\usepackage{mathrsfs}%
\usepackage[title]{appendix}%
\usepackage{xcolor}%
\usepackage{textcomp}%
\usepackage{manyfoot}%
\usepackage{booktabs}%
\usepackage{algorithm}%
\usepackage{algorithmicx}%
\usepackage{algpseudocode}%
\usepackage{listings}%
\usepackage{lineno}

\graphicspath{{./}}

\begin{document}

\title[Article Title]{Sub-unit cell engineering of CrVO$_3$ superlattice thin films}

\author*[1]{\fnm{Claudio} \sur{Bellani}}\email{claudio.bellani@kuleuven.be}\equalcont{These authors contributed equally to this work.}

\author[2,3 ]{\fnm{Simon} \sur{Mellaerts}}\email{simon.mellaerts@kuleuven.be}
\equalcont{These authors contributed equally to this work.}

\author[2]{\fnm{Wei-Fan} \sur{Hsu}}
\author[2]{\fnm{Koen} \sur{Schouteden}}
\author[2]{\fnm{Alberto} \sur{Binetti}}
\author[4]{\fnm{Arno} \sur{Annys}}
\author[4]{\fnm{Zezhong} \sur{Zhang}}
\author[4]{\fnm{Nicolas} \sur{Gauquelin}}
\author[4]{\fnm{Johan} \sur{Verbeeck}}
\author[5]{\fnm{Jesús} \sur{López-Sánchez}}
\author[5]{\fnm{Adolfo} \sur{del Campo}}
\author[6]{\fnm{Soon-Gil} \sur{Jung}}
\author[7]{\fnm{Tuson} \sur{Park}}
\author[2,8]{\fnm{Michel} \sur{Houssa}}
\author[2]{\fnm{Jean-Pierre} \sur{Locquet}}
\author*[1]{\fnm{Jin Won} \sur{Seo}}\email{maria.seo@kuleuven.be}

\affil*[1]{\orgdiv{Department of Materials Engineering}, \orgname{KU Leuven}, \orgaddress{\city{Leuven}, \postcode{3001}, \country{Belgium}}}

\affil[2]{\orgdiv{Department of Physics and Astronomy}, \orgname{KU Leuven},  \orgaddress{\city{Leuven}, \postcode{3001}, \country{Belgium}}}

\affil[3]{\orgdiv{Department of Electrical Engineering and Information Technology}, \orgname{ETH Zurich},  \orgaddress{\city{Zurich}, \postcode{8092}, \country{Switzerland}}}

\affil[4]{\orgdiv{Electron Microscopy for Materials Research (EMAT)}, \orgname{University of Antwerp},  \orgaddress{\city{Antwerp}, \postcode{2020}, \country{Belgium}}} 

\affil[5]{\orgdiv{Department of Electroceramics}, \orgname{Instituto de Cerámica y Vidrio - Consejo Superior de Investigaciones Cientificas (ICV-CSIC)},  \orgaddress{\city{Madrid}, \postcode{28049}, \country{Spain}}} 

\affil[6]{\orgdiv{Department of Physics Education}, \orgname{Sunchon National University},  \orgaddress{\city{Suncheon}, \postcode{57922}, \country{Republic of Korea}}} 

\affil[7]{\orgdiv{Center for Extreme Quantum Matter and Functionality (CeQMF) and Department of Physics}, \orgname{Sungkyunkwan University},  \orgaddress{\city{Suwon}, \postcode{16419}, \country{Republic of Korea}}} 

\affil[8]{\orgname{Imec},  \orgaddress{\city{Leuven}, \postcode{3001}, \country{Belgium}}}

\abstract{Ordered corundum oxides introduce new prospects in the field of functional oxides thin films, complementing the more widely studied class of ABO$_3$ perovskites. In this work, we take advantage of the layer-by-layer growth regime to fabricate epitaxial CrVO$_3$ superlattice thin films with atomic-scale accuracy on the periodic arrangement of Cr and V layers. By means of X-ray diffraction, scanning transmission electron microscopy and Raman spectroscopy, we confirm the thickness control in the sub-unit cell scale, alternating 3, 2 or 1 single atomic layers of Cr$_2$O$_3$ and V$_2$O$_3$. For the first time, we stabilize the ilmenite phase of CrVO$_3$ (space group R\(\bar{3}\)) and compare the functional properties of the thin film with those calculated by density functional theory. This novel approach to the growth of ordered corundum oxides opens the path towards the stabilization of new complex oxides with tailored properties by varying the composition and the superlattice period, ultimately broadening the family of functional rhombohedral oxides.}

\maketitle

\section{Introduction}\label{sec1}
\begin{figure*}
    \centering
    \includegraphics[width=0.8\linewidth]{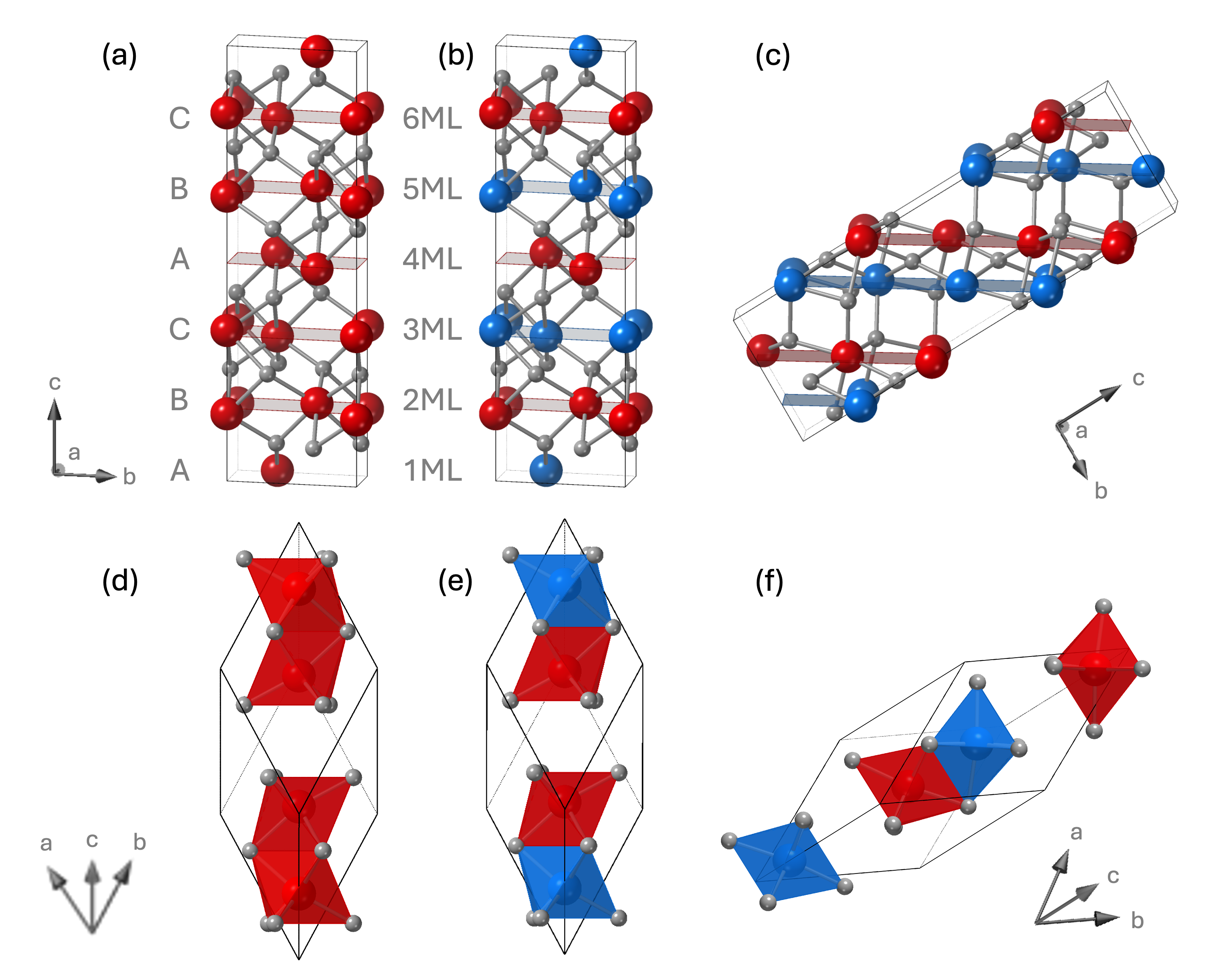}
    \caption{Crystal structures of the ordered corundums. Vanadium, chromium and oxygen atoms are represented in red, blue and grey, respectively. a,d) Conventional and primitive corundum unit cell (R\(\bar{3}\)c). The 6 atomic layers are highlighted as well as the ABC stacking. b,e) conventional and primitive ilmenite unit cell (R\(\bar{3}\)). c,f) Conventional and primitive LNO unit cell (R3c).}
    \label{fig:crystalmaker}
\end{figure*}

The field of complex oxides has been largely dominated by perovskite oxides, which are among the most extensively studied ABO$_3$ phases. Their widespread interest stems from their diverse functional properties, including ferroelectricity, colossal magnetoresistance, multiferroicity, and metal-insulator transitions. This versatility arises from their inherent instability to energy-lowering structural distortions, ultimately governed by the imbalance of ionic radii at the A and B sites, which is quantified by the Goldschmidt tolerance factor~\cite{goldschmidt1926gesetze,Glazer1972}. In nearly half of all perovskites, this instability manifests as an antiferrodistortive octahedral rotation pattern, $a^-a^-c^+$, resulting in an orthorhombic phase with space group Pbnm~\cite{chen2018energetics,Xiang2017}. Interestingly, hexagonal oxides exhibit similar tolerance factors and often compete with orthorhombic polymorphs, as for example in the nickelates~\cite{Catalano_2018}. However, when the ionic radii of the A and B cations become more comparable, i.e., $r_\text{ionic}$(A) $\approx$ $r_\text{ionic}$(B), the system transitions towards the regime of ordered corundum compounds~\cite{Mundy2022}. These parent crystal structure is illustrated in Figure~\ref{fig:crystalmaker}a and \ref{fig:crystalmaker}d with R\(\bar{3}\)c space group, of which Al$_2$O$_3$, Cr$_2$O$_3$ and V$_2$O$_3$ are most prominent examples. By defining A and B to be non-equivalent and ordered in the unit cell, the ilmenite (R\(\bar{3}\)) and polar LNO (R3c) symmetries can be established, as shown in Figure~\ref{fig:crystalmaker}b and \ref{fig:crystalmaker}c, respectively. These layered corundum compounds differ both structurally and chemically from the class of perovskites. Their small tolerance factor carries profound implications. First, they exhibit a high crystal density, where both A and B cations reside in equivalent octahedral environments. Second, the presence of face-sharing octahedra results in extremely short A-B distances, facilitating direct metal-metal bonding—distinct from the hybridization of oxygen atoms in A-O-B bonds found in perovskite materials. Lastly, the A and B sites can be occupied by similar transition metals, giving rise to strong electron correlations and associated phenomena, such as superconductivity and metal-insulator transitions (MITs)~\cite{Imada1998}.

In light of these considerations, the stabilization of new ordered corundum thin films offers intriguing opportunities in the field of functional oxides. On the one hand, the combination of two magnetic sublattices with strong electron correlations can lead to MITs; on the other hand new multiferroic compounds can be stabilized if ordered in the polar space group R3c~\cite{Cai2017,zhao2019first}. All together, polar metals with magnetoelectric coupling can be targeted~\cite{Zhou_2020,Spaldin2023}. To date, ordered corundum materials have been primarily prepared as powders by solid-state~\cite{chamberland1970preparation,syono1971high,sheikh1993phase,sohn1994microwave,aimi2011high,li2012polar,akaogi2015high,rodrigues2020unveiling} or wet-chemical synthesis~\cite{swoboda1958new,baraton1994vibrational,meena2020synthesis} methods, and as thin films starting from mixed precursors~\cite{popova2008systematic,varga2012epitaxial,varga2018controlling,miura2019growth}. As an alternative approach, we propose to leverage the atomic-layered growth via oxygen-assisted Molecular Beam Epitaxy (MBE) to enable sub-unit cell engineering of corundum compounds and the stabilization of new ordered corundum phases. This method provides several key advantages: access to a wider variety of A-B cation combinations, the opportunity to design tailor-made superlattice configurations, and the possibility of device integration. 

In this work, we experimentally demonstrate the feasibility of constructing Cr$_2$O$_3$/V$_2$O$_3$ superlattices with sub-unit-cell stacking. As the ultimate realization of this approach, we stabilize the ilmenite phase of CrVO$_3$ for the first time, representing the most extreme case: a 1-monolayer superlattice. The ordering of the alternating layers in this sub-unit cell regime has been directly evidenced by scanning transmission electron microscopy (STEM) combined with energy dispersive X-ray spectroscopy (EDX) and electron energy loss spectroscopy (EELS). Additionally, Raman spectroscopy has been employed to confirm symmetry lowering in these superlattices, as well as to verify the local ordering of the 1~ML CrVO$_3$ superlattice into the R\(\bar{3}\) symmetry of the ilmenite structure.
Finally, first-principles calculations predict that CrVO$_3$ is a ferromagnetic insulator, consistent with previous studies~\cite{le2018properties, zhao2019first}.

\section{Methods}\label{sec:methods}
\begin{figure*}
    \centering
    \includegraphics[width=0.78\linewidth]{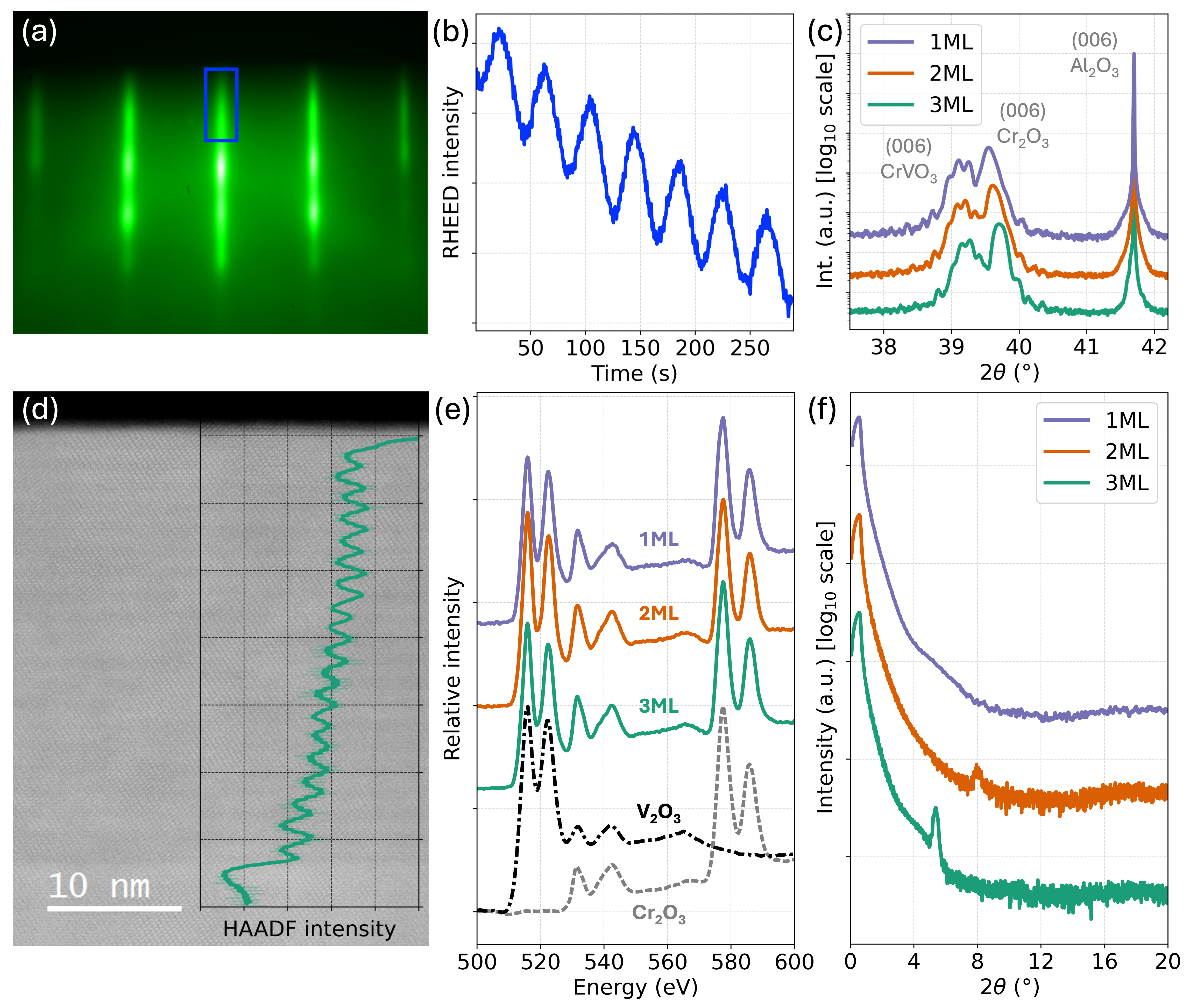}
    \caption{Structural and EELS characterization of the superlattice thin films. a) RHEED pattern at the end of the growth of the Cr$_2$O$_3$ buffer layer along the \(<1\bar{2}10>\) direction. The blue rectangle indicates the region where the intensity has been recorded as a function of time. b) RHEED intensity as a function of the deposition time. Each periodic oscillation corresponds to the completion of a single atomic layer. c) HRXRD of the (006) peak of the corundum substrate, buffer layer and superlattice thin films. d) Cross-sectional HAADF-STEM image of the 3~ML superlattice thin film along the \(<1\bar{2}10>\) ZA. The interface with the Cr$_2$O$_3$ buffer layer is visible in the bottom part of the image. The inset plot represents the HAADF intensity along the z-direction and averaged across the width of the image. e) STEM-EELS spectra near the V L-edge, O K-edge and Cr L-edge. The V L3-edge is used to align and normalize the intensity of the spectra. The Cr$_2$O$_3$ and V$_2$O$_3$ references were obtained from epitaxial thin films grown in the same MBE system. f) XRR of the series of superlattice thin films.}
    \label{fig:xrd-haadf-eels-xrr}
\end{figure*}

CrVO$_3$ thin films were deposited by oxygen-assisted MBE on 1x1~cm Al$_2$O$_3$-(0001) substrates at 700~°C in ultra-high vacuum (chamber base pressure \( 10^{-10} \)~mbar). The molecular oxygen partial pressure was kept constant at \( 2\cdot10^{-6} \)~mbar during the deposition, as measured with a residual gas analyzer connected to the growth chamber. The in-plane lattice mismatch of bulk V$_2$O$_3$ and Cr$_2$O$_3$ with the substrate is 4.2 \% and 4.3 \% respectively. Therefore a Cr$_2$O$_3$ buffer layer was always deposited first on the Al$_2$O$_3$ surface and followed by fully relaxed CrVO$_3$ thin films~\cite{dillemans2014evidence}. The fluxes of evaporated Cr and V were calibrated with a quartz crystal microbalance at 0.03~\text{\AA}/s, while reflection high-energy electron diffraction (RHEED) was used to monitor in situ the layer-by-layer epitaxial growth of the crystalline oxide films. The superlattice structures were obtained by alternating the opening and closing of the Cr and V cells shutters. The timing of the shutters switching was calibrated to achieve the desired periodicity, using the RHEED oscillations characteristic of a layer-by-layer growth. The calibration was based on the RHEED oscillations monitored during the deposition of the Cr$_2$O$_3$ buffer layer (Figure \ref{fig:xrd-haadf-eels-xrr}a-b) and kept constant throughout the entire growth of the thin film. The targeted structures consist of a sequence of the following repetition units: 3V+3Cr atomic layers, 2V+2Cr atomic layers, and 1V+1Cr atomic layers. Finally, a CrVO$_3$ thin film with a 1:1 Cr:V stoichiometry was grown by co-deposition of Cr and V as a benchmark for the superlattice structures. Simple Cr$_2$O$_3$ and V$_2$O$_3$ epitaxial thin films on sapphire are used as a reference~\cite{Hsu2024}.

The epitaxial growth was confirmed by X-ray diffraction (XRD) and STEM, while the accuracy of the sample layering was evaluated by X-ray reflectivity (XRR), analytical STEM and Raman spectroscopy. The XRD and XRR experiments were conducted with a Rigaku TF-XRD-300 system, equipped with a dual rotating Cu anode and a HyPix-3000 detector. The system was operated with a Ge220x2 monochromator in the incident optics for the XRD measurements.

Raman spectra were acquired at room temperature using a 100× objective with a numerical aperture of 0.95. An output laser power of 40~mW was used to avoid the overheating effects and possible damage to the sample. Raman spectra were collected in the spectral range 65-1160~cm$^{-1}$ using an 1800~g~mm$^{-1}$ grating. Under optimal conditions, the spectral resolution is 0.02~cm$^{-1}$. In turn, the vibrational properties were also evaluated as a function of polarization angle. Here, the polarization analyzer detector is adjusted from 0° (parallel polarization, VV) to 90° (cross polarization, VH) against the incident polarization. Raw Raman data were processed with Witec Project Plus software (version 2.08).

Cross-sectional lamellae were prepared by focused ion beam (FIB) for STEM imaging and elemental mapping using DualBeam Nova 600 NanoLab, then analyzed with a cold-FEG JEOL ARM200F microscope operated at 200~kV, equipped with a STEM-Cs corrector, a Centurio EDX spectrometer and a Gatan Tridiem electron energy loss spectrometer. EELS spectra were collected with a collection semi-angle of 16.7~mrad and a beam current of 120~pA. For the 1~ML and 2~ML samples, thinner lamellas were prepared on a Dual beam FIB/SEM Helios Nanolab 650. A more detailed analysis was performed on a Themis Z microscope, which is an aberration-corrected STEM, Thermo Fisher Scientific Titan3 60-300 operated at an acceleration voltage of 200~kV, equipped with a high-brightness field-emission electron source (X-FEG) and an electron monochromator excited to 1.4 providing an energy resolution better than 200~meV while keeping a spatial resolution of 1~Å. EDX and EELS were acquired simultaneously using a collection semi-angle of 56~mrad and a beam current of 50~pA. Multiple EELS data were acquired on a prototype Iliad EELS Spectrometer with Zebra EELS detector developed by Thermo Fisher with a dispersion of 0.05~eV/pixel~\cite{jannis2024multi}. Core-loss EELS data analysis was performed using model-based quantification with the pyEELSmodel software~\cite{pyEELSMODEL}. The core-loss data was aligned using the zero-loss peak (ZLP) and spatially binned to improve signal-to-noise ratio (SNR) of the spectra and facilitate the fitting procedure. Relative estimates of spatial variations of Cr and V can be extracted using model-based fitting reliably. The model consisted of three parts. First, a convexity constrained linear background~\cite{van2023convexity}. Second, core-loss edges using the cross sections for O, V and Cr~\cite{zhang2025relativistic}. Third, a component considering multiple scattering based on the low loss signal. For EDX elemental mapping of the 2~ML and 1~ML superlattices, multi-frame EDX data was acquired, using the simultaneously acquired HAADF signal, all frames were aligned using the SuperAlign routine~\cite{superalign}.

The resistivity of the 1~ML superlattice thin film was measured in the Van der Pauw configuration with a Keithley 4200-SCS parameter analyzer between 50 and 300~K. UV-Vis absorbance spectra were acquired using a Tecan Infinite 200 PRO microplate reader. A magnetic property measurement system (MPMS) based on a superconducting quantum interference device (SQUID) magnetometer was used to characterize the magnetic properties of a selection of thin films. The relatively strong diamagnetic signal due to the substrate was fitted in the the high fields regions (5~kOe $<|H|<$~30 kOe) and subsequently subtracted from the measured magnetization (see Figures~S1 and S2).

All density functional theory (DFT) calculations were carried out using the Vienna Ab initio Simulation Package (VASP)~\cite{VASP}. The interactions between electrons and ions were treated using the projector augmented-wave (PAW) method~\cite{PAWmethod}, and the electronic wave functions were expanded with a plane-wave cutoff energy of 600~eV. The PBEsol exchange-correlation functional~\cite{PBEsol} was employed throughout, while the localized nature
of the $3d$ electrons of the cations was treated by a Hubbard correction $U$ within the rotationally invariant approach proposed by Dudarev et al.~\cite{Dudarev}. Structural relaxations were performed with a force convergence criterion of 0.005~eV/Å, and the Brillouin zone was sampled using a $\Gamma$-centered $12\times12\times12$ $k$-point mesh. For the electronic and magnetic calculations, a dense $24\times24\times24$ mesh was used with an energy convergence criterion of $10^{-6}$~eV. The phonon dispersions were evaluated by density functional
perturbation theory (DFPT) and with the use of the PHONOPY package~\cite{phonopy}.

\section{Results}\label{sec:results}

\subsection{Sub-unitcell engineered \texorpdfstring{CrVO$_3$}{CrVO3}}
\begin{figure*}
    \centering
    \includegraphics[width=0.8814\linewidth]{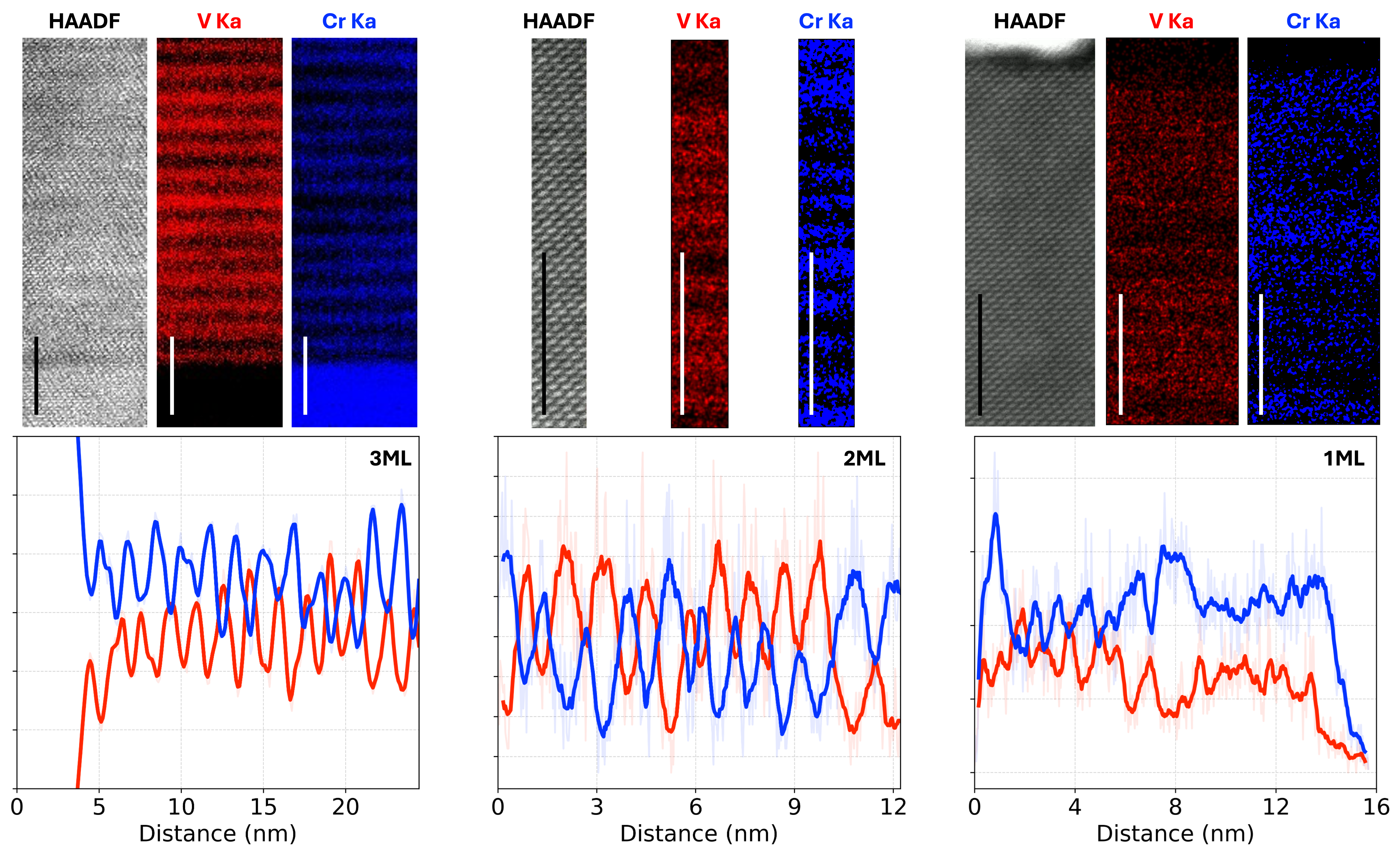}
    \caption{Cross-sectional HAADF-STEM (\(<1\bar{2}10>\) ZA) images and corresponding EDX chemical maps of the 3~ML, 2~ML, and 1~ML superlattice thin films (from left to right). V and Cr distributions are mapped using the intensity of their respective K\(\alpha\) lines. The 3~ML sample map was processed with a Wiener filter, while the corresponding intensity profile presents the unfiltered data. At the bottom, a portion of the Cr$_2$O$_3$ buffer layer is also visible. To minimize the impact of surface steps on the chemical maps, the 2~ML and 1~ML samples were imaged from FIB lamellae thinner than 20~nm. Given the short interlayer distances of the 2~ML and 1~ML films, multi-frame imaging and spectroscopy data were acquired and aligned to correct for sample drift and scanning distortions. All scalebars represent 5~nm.}
    \label{fig:edx}
\end{figure*}
Figure ~\ref{fig:xrd-haadf-eels-xrr}a shows the RHEED pattern at the end of the growth of a Cr$_2$O$_3$ buffer layer, confirming the epitaxial growth of the film. The RHEED intensity can be recorded as a function of time as represented in Figure ~\ref{fig:xrd-haadf-eels-xrr}b: the measured periodic oscillations indicate the layer-by-layer growth regime of the corundum thin film, and the period of about 40 s corresponds to the time required to form a complete single atomic layer at the calibrated growth rate. 

The epitaxial growth of the buffer layer and of the superlattice thin film on the sapphire substrate is also confirmed by high-resolution XRD pattern and and high-angle annular dark-field scanning transmission electron microscopy (HAADF-STEM) images in Figure~\ref{fig:xrd-haadf-eels-xrr}c-d. Additionally, the STEM-EELS spectra, measured across the entire thickness of the films, confirm the oxidation states of Cr$^{3+}$ and V$^{3+}$. This conclusion is supported by direct comparison with the reference spectra shown in Figure~\ref{fig:xrd-haadf-eels-xrr}e, as well as with previously reported results in the literature~\cite{lin1993valence,hebert2002oxygen,asa2018interdiffusion}. It is also noteworthy that the buffer layer exhibits the Cr$^{3+}$ oxidation state. The fact that both Cr and V are stabilized in the $3+$ oxidation state is particularly significant for two reasons. First, it rules out other common vanadium oxidation states, such as V$^{4+}$ and V$^{5+}$~\cite{kosuge1967phase,katzke2003theory}. Second, it marks a difference with existing \(\text{M}^{2+}\text{TiO}_3 \) and \(\text{M}^{2+}\text{MnO}_3 \) ilmenites where two metals are arranged in a layered structure with 2+ and 4+ oxidation states, respectively~\cite{ishikawa1956study,chamberland1970preparation,sohn1994microwave,akaogi2015high}. Possibly, this can be attributed to the high stability of the Cr$^{3+}$ cation~\cite{white1976system} in comparison to Cr$^{2+}$ and Cr$^{4+}$. As a result, V$^{3+}$ seems to be stabilized when confined between Cr$_2$O$_3$ layers. A logical extension would be to examine vanadium in combination with elements that are readily oxidized to $M^{2+}$ and/or $M^{4+}$ states.

Compared to the Cr$_2$O$_3$ buffer layer, the superlattice thin film exhibits a distinct oscillation of the HAADF signal along the out-of-plane direction, attributed to the different atomic number (Z) contrast between Cr and V. However, such contrast remains weak to effectively highlight and measure the periodicity of Cr-V superstructures with shorter periods (see Figure~S3) because of the small Z difference.

In addition to the (0006) and (00012) reflections that emerge from the ABC stacking in the corundum structure with space group R\(\bar{3}\)c, additional diffraction occurs in the 3~ML and 2~ML superlattices due to the artificially introduced periodicity. This results into higher-order reflections in the XRR (see Figure~\ref{fig:xrd-haadf-eels-xrr}f) with a d-spacing of \(d_\text{3~ML} = 16.5 \pm 0.4 \, \text{Å}\) and \(d_\text{2~ML} = 11.0 \pm 0.4 \, \text{Å}\), respectively. On the other hand, the removal of the glide reflection in the 1~ML superlattice with expected R\(\bar{3}\) space group symmetry would lead to a (0003) reflection in the diffractogram at $2\theta\approx$19° corresponding to \(d_\text{1~ML} = 4.7 \text{Å}\). However, such a reflection proved to be hard to detect, possibly due to the background and noise levels.

\begin{figure*}
    \centering
    \includegraphics[width=\linewidth]{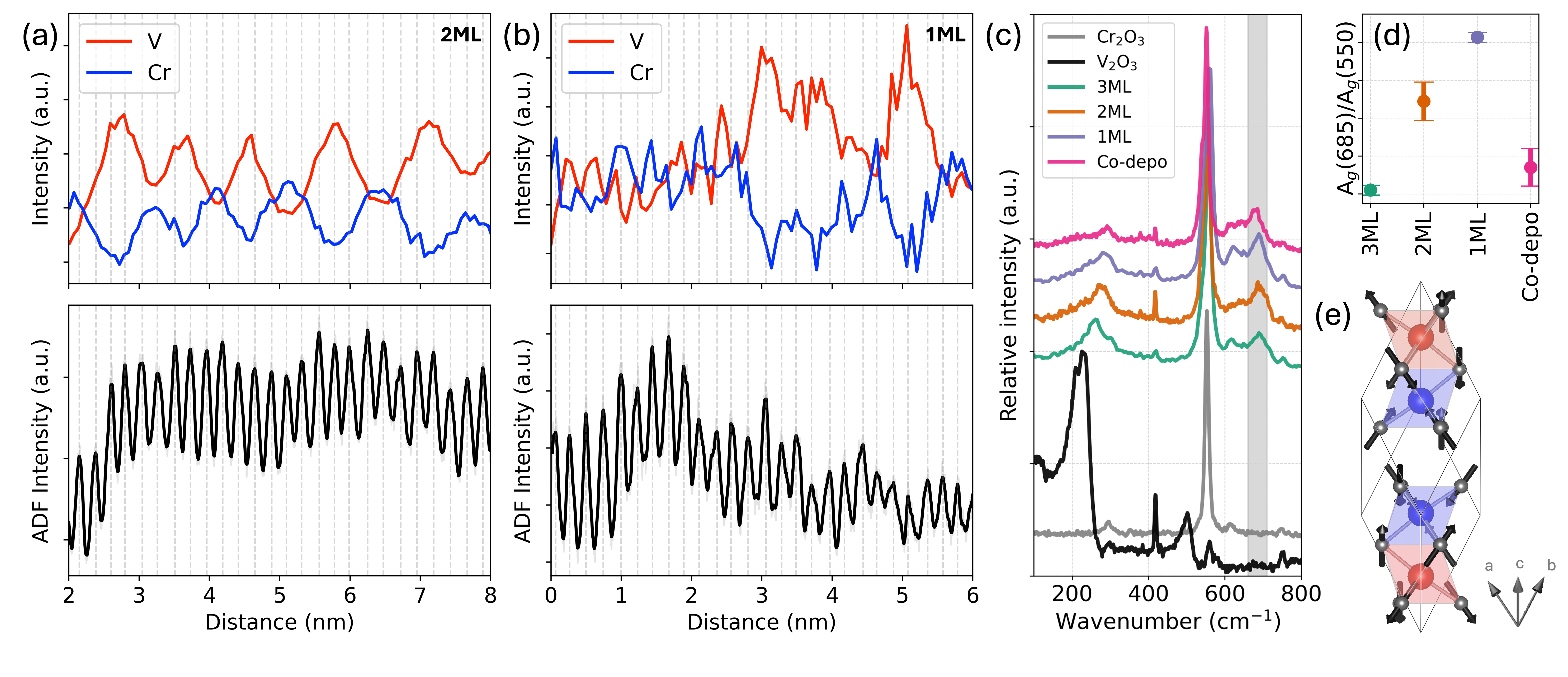}
    \caption{STEM-EELS and Raman measurements confirm the periodicity of the superlattice thin films. a-b) HAADF-STEM and EELS profiles of the 2~ML and 1~ML superlattices. Vertical dashed lines are drawn matching the peaks in the HAADF signal, each corresponding to a single layer of cations. c) Raman spectra of the superlattice thin films, compared to pure V$_2$O$_3$, Cr$_2$O$_3$, and a co-deposited CrVO$_3$ film. All the V$_2$O$_3$ and CrVO$_3$ films are grown on a Cr$_2$O$_3$ buffer layer. As discussed in the text, an additional A\(_g\) mode appears in the CrVO$_3$ samples around 685~cm\(^{-1}\). The intensity of this mode is normalized to the intensity of the A\(_{1g}\) mode at 550 cm\(^{-1}\) and is shown in the inset plot. The error bars represent the standard deviation of the signal noise. d) Illustration of the Raman activated A\(_{g}\) mode. The expansion (contraction) of the red (blue) octahedra is represented by the DFT-calculated eigenvectors (black arrows).}
    \label{fig:EELSandRaman}
\end{figure*}

To confirm the measured periodicity and evaluate the interface quality of the periodical superstructures, STEM-EDX was performed. Figure~\ref{fig:edx}) includes the elemental maps of all three superlattice thin films, as well as the corresponding line profile calculated along the out-of-plane direction. The Cr and V signals are periodic, with the Cr maxima matching with the V minima. The period of the superlattice was calculated as the average distance between the Cr and V maxima, including the standard deviation to quantify its variability. In particular, the periods \(d_\text{3~ML}= 16.3 \pm 0.9 \, \text{Å}\) and \(d_\text{2~ML} = 11.0 \pm 1.2 \, \text{Å}\) are found to be in agreement with XRR. On the other hand, the 1~ML film presents a Cr-V periodicity only locally, \(d_\text{1~ML} = 7.9 \pm 0.7 \, \text{Å}\), explaining why the (0003) reflection was harder to detect in the X-ray diffractograms. Elemental mapping was also performed by STEM-EELS on the 2~ML and 1~ML samples. An absolute quantitative analysis is challenging due to the overlap of the V, Cr, and O edges (see Figure~\ref{fig:xrd-haadf-eels-xrr}e), but a relative estimate of the spatial distribution of Cr and V can be obtained by model-based quantification with the pyEELSmodel software~\cite{pyEELSMODEL}. The resulting spatial distributions of Cr and V along the out-of-plane direction in the 2~ML and 1~ML superlattices are plotted in Figure~\ref{fig:EELSandRaman}a-b together with the HAADF signal. STEM-EELS confirms the ordering of the 2~ML superlattice (period \(d_\text{2~ML} = 10.9 \pm 1.3 \, \text{Å}\)) and the local cation ordering in the 1~ML thin film. However, the alignment of the Cr-V spatial distribution with the HAADF signal also highlights the errors in the 2Cr+2V layers stacking; e.g. the Cr (or V) peak occasionally spans over 3 HAADF peaks corresponding to 3 atomic layers. The deviations from the designed Cr-V stacking periodicity can be explained by extrinsic factors, such as instabilities of the Cr and V evaporation fluxes during the deposition. However, intrinsic factors like interdiffusion cannot be ruled out and they can impact the quality of the superlattice interfaces. Notably, antisite defects are expected to be structurally more favorable in ABO$_3$ ordered corundum structures than perovskites because of the similar oxygen environments for A and B cations. Both sites have a six-fold oxygen coordination in a distorted octahedra, while in the perovskite lattice A and B are surrounded by 12 and 6 O atoms, respectively. In the specific case of CrVO$_3$, we have shown that Cr and V have also the same valence state (Figure~\ref{fig:xrd-haadf-eels-xrr}e). In addition, Zhao et al.~\cite{zhao2019first} calculated that the properties of ordered ABO$_3$ corundums are not affected by the swap of A and B cations, suggesting a limited impact of antisite defects.

\subsection{Phonon spectroscopy to probe symmetry breaking}

The introduced local symmetry breaking in the superlattices was evaluated by polarized Raman spectroscopy. As shown in Figure~\ref{fig:EELSandRaman}, an additional band is detected around 685~cm\(^{-1}\) in the CrVO$_3$ samples in comparison to the sesquioxides Cr$_2$O$_3$ and V$_2$O$_3$~\cite{Hsu2024,Mellaerts2024}. This is the highest observed frequency mode in the range 100-1150 cm$^{-1}$ and its intensity increases as the superlattice period is reduced (Figure~\ref{fig:EELSandRaman}d). A similar observation was made by Baraton et al.~\cite{baraton1994vibrational} for FeCrO\(_{3}\) powders obtained by co-precipitation, where they identified a new peak, the highest-frequency phonon mode around 700 cm\(^{-1}\), which was not observed in the parent oxides. As noted by the authors, the presence of the extra Raman band is a proof of cation ordering. The additional band is assigned to the inactive A\(_{2g}\) mode of \(\text{A}_2\text{O}_3\) corundum (R\(\bar{3}\)c space group) which becomes the Raman active A\(_{g}\) mode in the \(\text{ABO}_3\) ilmenite structure (R\(\bar{3}\) space group). Both modes correspond to the contraction and expansion of AO\(_6\) and BO\(_6\) octahedra in successive {0001} planes (Figure~\ref{fig:EELSandRaman}e), resulting in the inactive A\(_{2g}\) if A\(=\)B in the parent \(\text{A}_2\text{O}_3\) corundum oxide, but Raman active (A\(_{g}\)) if A\(\neq\)B in the \(\text{ABO}_3\) ilmenite structure.

The same A\(_{g}\) assignment was made by Wang et al.~\cite{wang2008assignment} for MgTiO\(_{3}\): 10 Raman modes are observed in the Raman spectrum as predicted by group theory and match the DFT calculated frequencies for an ilmenite structure. The A\(_{g}\) mode is detected at 715~cm\(^{-1}\). Similarly, we calculated the frequencies of the Raman active modes in ilmenite CrVO$_3$ (Figure~S4): the calculated highest frequency mode is the A\(_{g}\) vibration described above and we assign it to the highest observed frequency mode at 685~cm\(^{-1}\). To further confirm the A\(_{g}\) assignment, the Raman spectra were also collected as a function of the polarization angle (Figure~S5). Consistent with the selection rules for the A\(_{g}\) mode, the band intensity exhibits a pronounced and gradual decrease when transitioning from parallel (VV) to cross (VH) polarization, providing a more robust evidence for the mode assignment.

Two additional A\(_{2g}\) modes are expected to be Raman active in the ilmenite R\(\bar{3}\) symmetry with their expected frequencies in the ranges \(\sim\) 186 cm\(^{-1}\) and \(\sim\)395 cm\(^{-1}\) as predicted by DFT (Table S1). The first overlaps with the A\(_{1g}\)-derived A\(_{g}\) mode, giving shape to an asymmetric peak (Figure~S6). The second was not detected, either due to a smaller Raman polarizability or an overlap with the peaks of the buffer layer. Taking all the factors into account, we have demonstrated the presence of R\(\bar{3}\) symmetry features in the Raman spectra of all the superlattice thin films and the co-deposited CrVO$_{3}$.
The last question to address is why such features are not observed solely in the 1~ML sample, which is designed to adopt the ilmenite structure with R\(\bar{3}\) symmetry. It is plausible to argue that the appearance of these features is caused by the symmetry breaking at the interfaces between the Cr$_2$O$_3$ and V$_2$O$_3$ layers. The same activation of a Raman inactive mode has already been observed in perovskite-based superlattices by Seong et al.~\cite{jeong2022atomistic}. Such an interpretation would also be consistent with the increase in intensity of the A\(_{g}\) mode as the density of interfaces in the superstructures rises (i.e., with the decreasing period of the superlattice), as shown in Figure~\ref{fig:EELSandRaman}d. In the broader context, it indicates that the cations in the 1~ML sample exhibit at least local ordering, as previously demonstrated by STEM-EDX and STEM-EELS. Furthermore, it suggests the presence of a self-ordering mechanism in the co-deposited sample, though less effective than the periodic opening and closing of the shutters, as evidenced by the weaker \( A_{g} \) peak. This result should not be surprising for two reasons. First, the structural stability of ilmenite CrVO$_{3}$, as calculated in this work, aligns with previous literature findings~\cite{le2018properties,zhao2019first}. Second, ilmenite structures have already been successfully synthesized using methods such as solid-state reactions and co-precipitation, which do not impose superstructure periodicity and must therefore rely on a form of self-ordering~\cite{akaogi2015high,baraton1994vibrational,chamberland1970preparation,sohn1994microwave,cloud1958crystal,matthias1949ferroelectricity,rodrigues2020unveiling,swoboda1958new}.

\begin{figure*}
    \centering
    \includegraphics[width=0.685\linewidth]{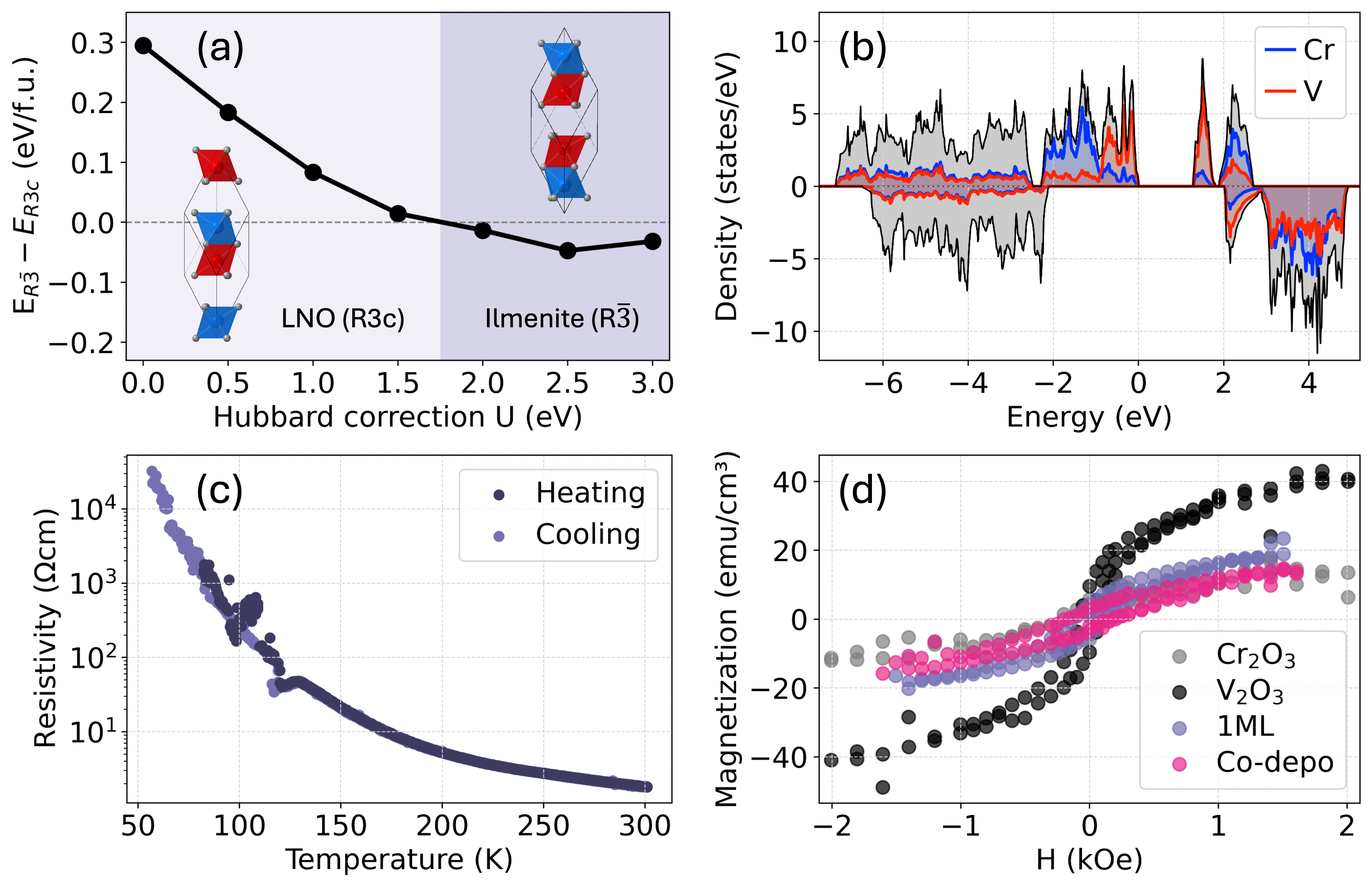}
    \caption{Functional properties of CrVO$_{3}$ with R$\bar{3}$ ilmenite structure. a) Energy comparison of the ilmenite (R$\bar{3}$) and the LNO (R3c) structures as a function of the Hubbard correction $U$ for CrVO$_{3}$. The background color and the crystal models indicate what structure is the most stable in the two regions of the phase diagram. b) The Cr and V contributions to the partial DOS of ilmenite CrVO$_{3}$. c) Resistivity of the CrVO$_{3}$ 1~ML superlattice thin film. d) Magnetization of CrVO$_{3}$, V$_2$O$_3$ and Cr$_2$O$_3$ thin films as a function of the external magnetic field H at 10~K and with H parallel to the in-plane lattice direction.}
    \label{fig:DFTandCo}
\end{figure*}

\subsection{Ferromagnetic insulating ilmenite}
By DFT calculations, the relative stability of both the non-centrosymmetric ilmenite and the polar LNO structure were evaluated as a function of Hubbard correction ($U$), as shown in Figure~\ref{fig:DFTandCo}a. It can be anticipated that a Hubbard value of $3$~eV is considered an appropriate approximation for CrVO$_{3}$, based on prior works of V$_2$O$_3$ and Cr$_2$O$_3$~\cite{Grieger2015,Shi2009}. Henceforth, it is inferred that the ilmenite structure is energetically favored. Subsequently, the electronic properties were evaluated, as depicted in Figure~\ref{fig:DFTandCo}a. An insulating behavior is observed with an estimated energy gap of $1.28$~eV. Finally, the magnetic ordering and main magnetic interactions were evaluated by considering five different short-period spin arrangements (see Figure~S7). By an energy comparison, it is shown that FM ordering is energetically favored above $2$~eV, with magnetic moments $M_V=2.02\mu_B$ and $M_{Cr}=3.07\mu_B$, in excellent agreement with prior work~\cite{zhao2019first}. Although a strong exchange interaction of $52.2$~meV exists along the short Cr-V dimer, the relative energy stability of FM is only $14$~meV/f.u. compared to the lowest AF state (at $U=3$~eV). This might suggest a relatively weak FM ordering.

Following up on the DFT calculations, a preliminary investigation into the functional properties of the 1~ML superlattice thin film was performed. The resistivity measurement as a function of temperature shows a clear insulating behavior as the sample is cooled down to 50K from room temperature (Figure ~\ref{fig:DFTandCo}c). The thin film is confirmed to be insulating by UV-VIS spectroscopy (see Figure~S8), but a reliable estimate of the band gap proved to be challenging due to the overlap with the optical features of the Cr$_2$O$_3$ buffer layer, i.e. its 3.1~eV band gap and crystal field excitations of Cr$^{3+}$ ions at 2.1 and 2.6~eV~\cite{sala2016resonant}. At last, the magnetic properties of the CrVO$_3$ 1~ML superlattice were assessed and compared to Cr$_2$O$_3$, V$_2$O$_3$ and co-deposited CrVO$_3$ thin films. The measured magnetization M as a function of the magnetic field H is shown in the Figure~\ref{fig:DFTandCo}d. All the samples present a weak ferromagnetic response along both field directions (in and out-of-plane) and at both probed temperatures (300~K and 10~K). The largest remnant magnetization is observed in the V$_2$O$_3$ thin film, in all the four combinations of H direction and sample temperature (see Figure~\ref{fig:DFTandCo}d and Figure~S9); this is in contrast with the expected paramagnetic behavior of V$_2$O$_3$~\cite{mcwhan1973metal}. Such weak ferromagnetism could arise from the presence of magnetic impurities. On the other hand, the CrVO$_3$ thin films exhibit a larger coercive field (H$_c\sim$85 Oe at 300~K and H$_c\sim$130 Oe at 10~K for the 1~ML superlattice), which could instead be explained by a partial magnetic ordering.

\section{Conclusions}\label{sec:conclusions}
 The results presented in this work introduce a novel approach to the synthesis of ordered corundum oxides, leveraging the atomic layer-by-layer growth of epitaxial thin films. CrVO$_{3}$ superlattice thin films have been grown by oxygen-assisted MBE, progressively reducing the thickness of the alternating Cr$_{2}$O$_{3}$ and V$_{2}$O$_{3}$ layers. Specifically, the M$_{2}$O$_{3}$ units consist of 3, 2 or 1 layer of cations, i.e. one half, one third or one sixth of the conventional corundum unit cell height. A detailed structural characterization showed how the superlattices can be controlled at the atomic scale, although the induced Cr-V alternation may be in competition with interdiffusion at the 1~ML limit at the growth temperature of 700°C. Moreover, antisite defects are expected to be structurally more favorable if compared to perovskites due to the similar sixfold O-coordination of the Cr and V sites. Despite these challenges, this novel approach towards the synthesis of ordered corundum structures opens a promising path for the growth of novel artificial functional oxide thin films, whose properties depend on the layering configuration in the corundum lattice and the chosen pair of cations. First-principles calculations have shown the potential of 1~ML limit, i.e. ilmenite and LNO structures, while the properties of sub-unit cell superlattices need to be explored in detail in future studies.

\section*
{Acknowledgments}\label{sec:acknowledgments}
This work was financially supported by the KUL C14/21/083, by the FWO infrastructure projects I000920N and AKUL/13/19, and the FWO IRI project I002123N. N. G., Z. Z., and J. V. acknowledge the funding from the European Union’s Horizon 2020 research and innovation program under Grant Agreement No. 823717–ESTEEM3. A. A. acknowledges the IMPRESS project that has received funding from HORIZON EUROPE framework program for research and innovation under grant agreement n. 101094299. J. L.-S. acknowledges the financial support from grant PID2023-151036OA-I00 funded by MICIU/AEI/10.13039/501100011033 and by ERDF, EU, and from grant RYC2022-035912-I funded by MCIU/AEI/10.13039/501100011033 and by the European Social Fund Plus (ESF+). This work was supported by National Research Foundation of Korea (NRF) grants funded by the Korean government (MSIT) RS-2023–00220471 and RS-2023–00240326.

\bibliography{sn-bibliography}

\clearpage
\appendix
\renewcommand{\thefigure}{S\arabic{figure}}
\setcounter{figure}{0}
\renewcommand{\thetable}{S\arabic{table}}
\setcounter{table}{0}

\section{Supplementary Material}

\begin{figure*}
    \centering
    \includegraphics[width=0.7\linewidth]{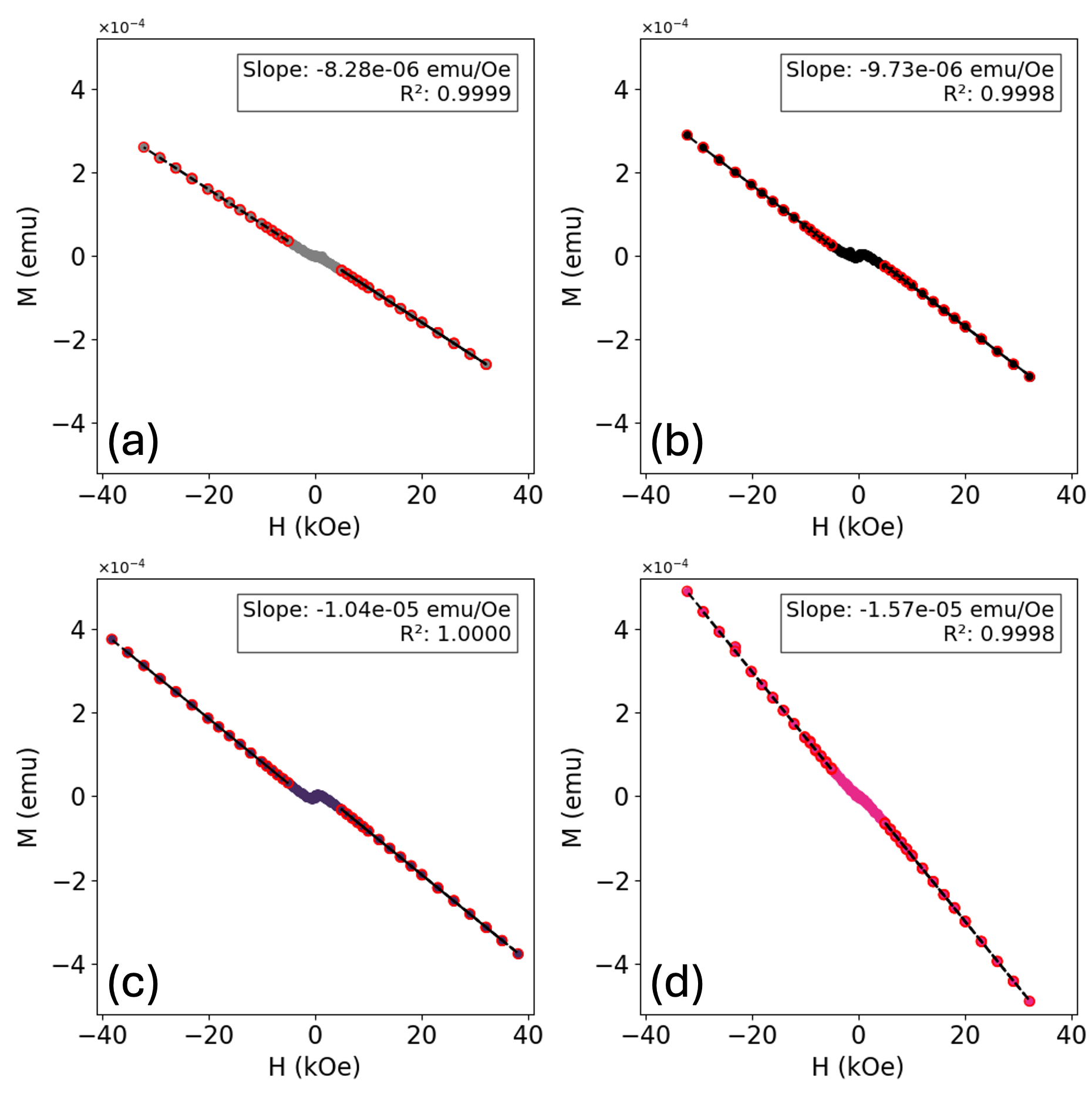}
    \caption{Magnetization of CrVO$_3$, Cr$_2$O$_3$, and V$_2$O$_3$ thin films as a function of the external magnetic field $H$ at 10~K, with the field applied parallel to the in-plane lattice direction. A linear fit in the high-field region (5~kOe~$<|H|<$~30~kOe) determines the diamagnetic component (negative slope). The data correspond to: (a) Cr$_2$O$_3$, (b) V$_2$O$_3$, (c) 1~ML CrVO$_3$, and (d) co-deposited CrVO$_3$. An equivalent procedure has been performed for all combinations of temperature (10~K and 300~K) and field direction (in-plane and out-of-plane).}
    \label{fig:highFields}
\end{figure*}

\begin{figure*}
    \centering
    \includegraphics[width=\linewidth]{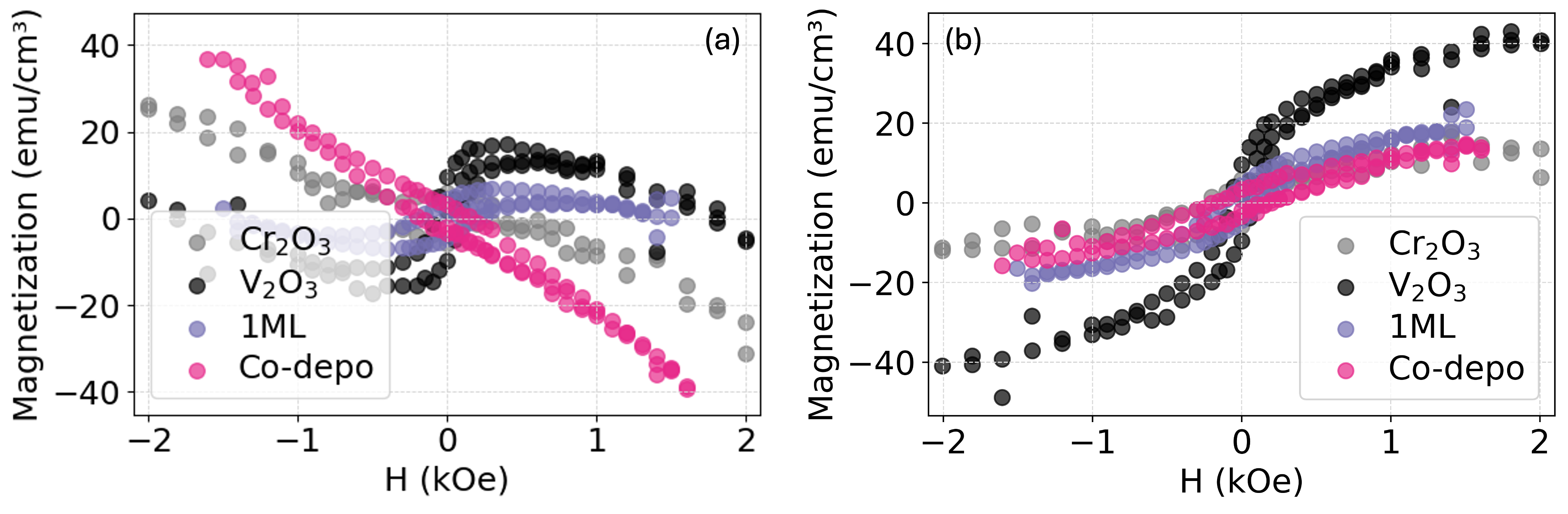}
    \caption{Magnetization of CrVO$_3$, Cr$_2$O$_3$, and V$_2$O$_3$ thin films as a function of the external magnetic field $H$ ($|H|<$~2~kOe) at 10~K, with $H$ applied parallel to the in-plane lattice direction. (a) Data normalized by the probed film volume, without subtraction of the diamagnetic contribution. (b) Data normalized by the probed film volume, with subtraction of the diamagnetic component using the slopes obtained from the linear fits shown in Figure~S1.}
    \label{fig:squidFit}
\end{figure*}

\begin{figure*}
    \centering
    \includegraphics[width=\linewidth]{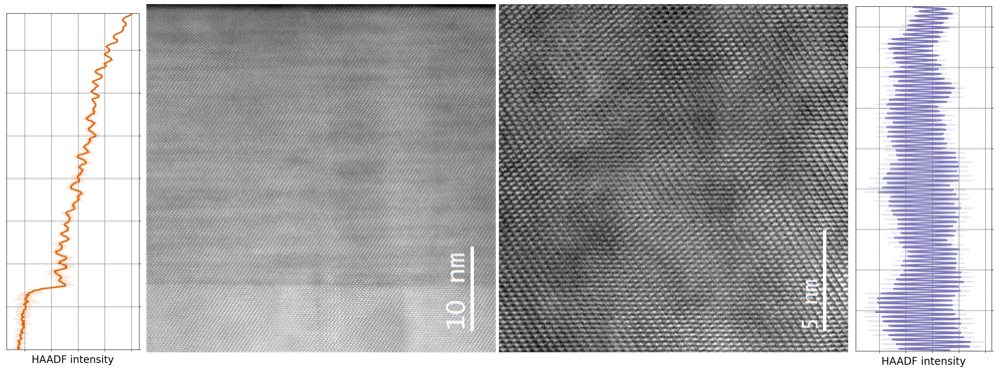}
    \caption{Cross-sectional HAADF-STEM image of the 2ML (left) and 1ML (right) superlattice thin films. The HAADF intensity is plotted along the out-of-plane direction and averaged across the width of the image. Differently from the 3ML case, it is not possible to distinguish the Cr and V layers due to the small Z-contrast.}
    \label{fig:moreHAADF}
\end{figure*}

\begin{table*}[h]
\centering
\begin{tabular}{|c|c|}
\hline
Original mode & Frequency (cm$^{-1}$)\\
\hline
$A_{2g}$ & 186 \\
\hline
$A_{1g}$ & 255 \\
\hline
$A_{2g}$ & 395 \\
\hline
$A_{1g}$ & 477 \\
\hline
$A_{2g}$ & 536 \\
\hline
\end{tabular}
\caption{All DFT-calculated $A_g$ modes of the ilmenite CrVO$_3$. The first column represents the original symmetry assignment with $R\bar{3}c$ for V$_2$O$_3$ and Cr$_2$O$_3$}
\end{table*}

\begin{figure*}
    \centering
    \includegraphics[width=0.7\linewidth]{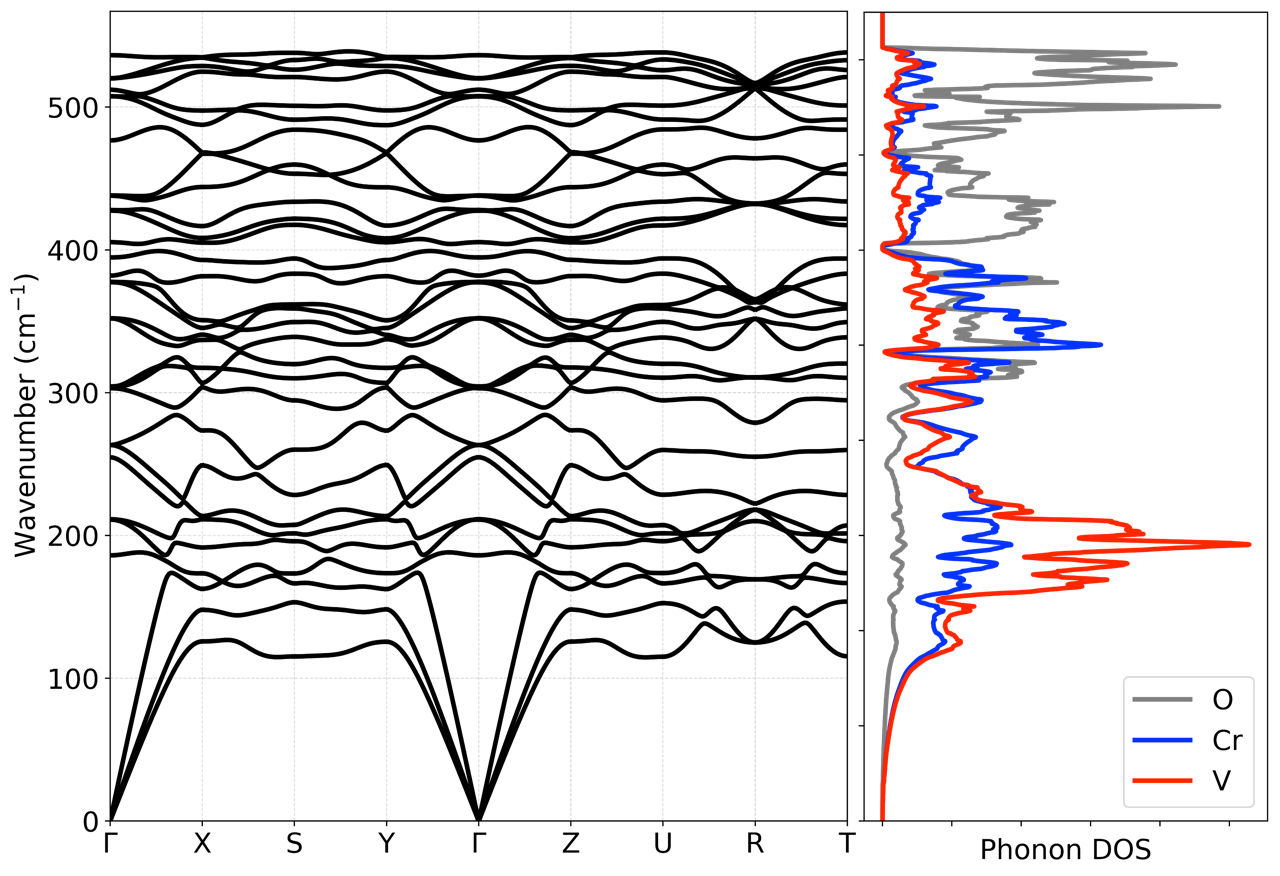}
    \caption{Calculated phonon spectrum of the ilmenite CrVO$_3$. On the right, the corresponding atomic-resolved phonon DOS.}
    \label{fig:phononDOS}
\end{figure*}

\begin{figure*}
    \centering
    \includegraphics[width=0.4\linewidth]{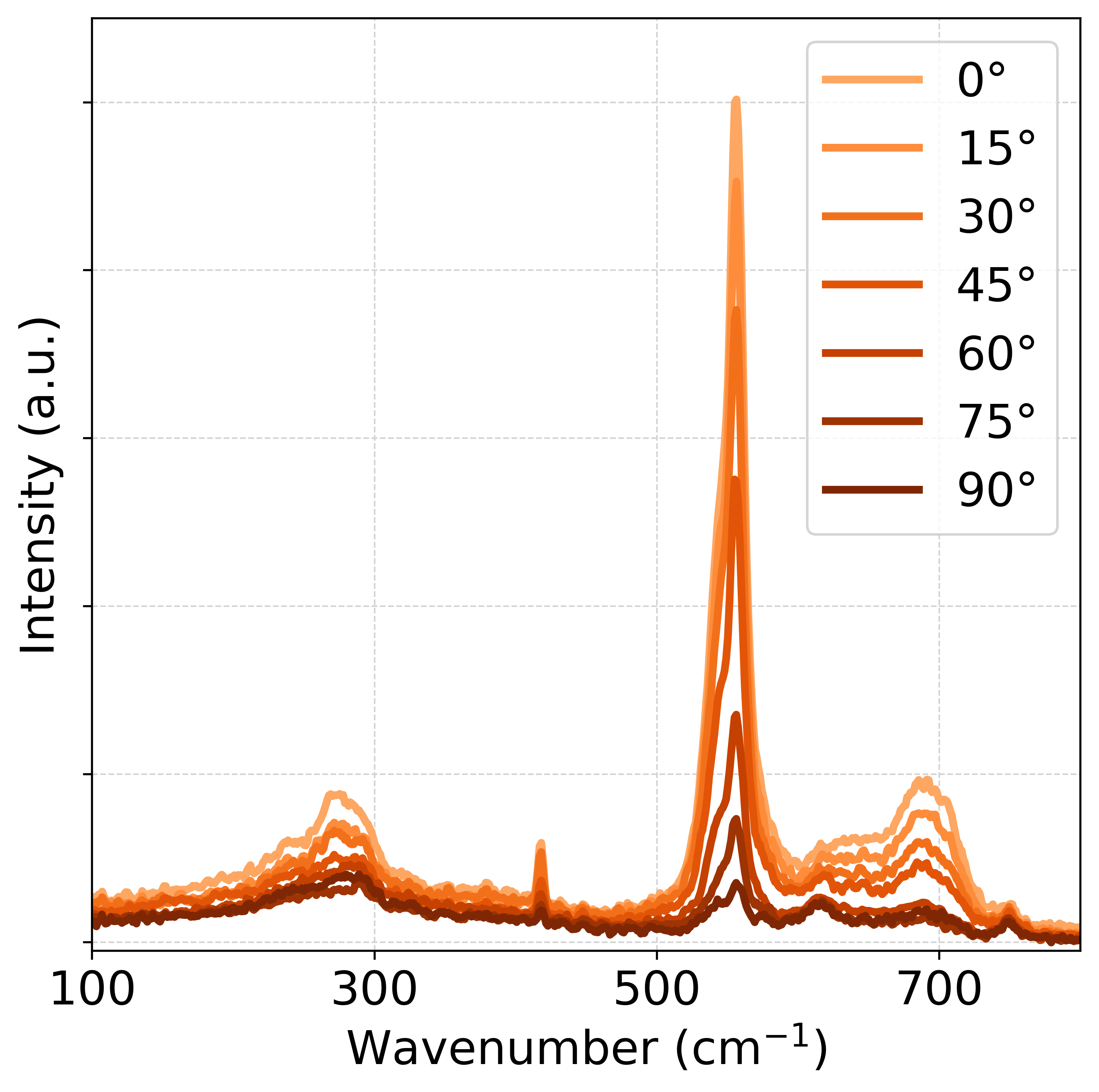}
    \caption{Raman spectra of the 2ML superlattice thin film as a function of the polarization angle, ranging between 0° (parallel polarization, VV) and 90° (cross polarization, VH).}
    \label{fig:vsPolarization}
\end{figure*}

\begin{figure*}
    \centering
    \includegraphics[width=0.4\linewidth]{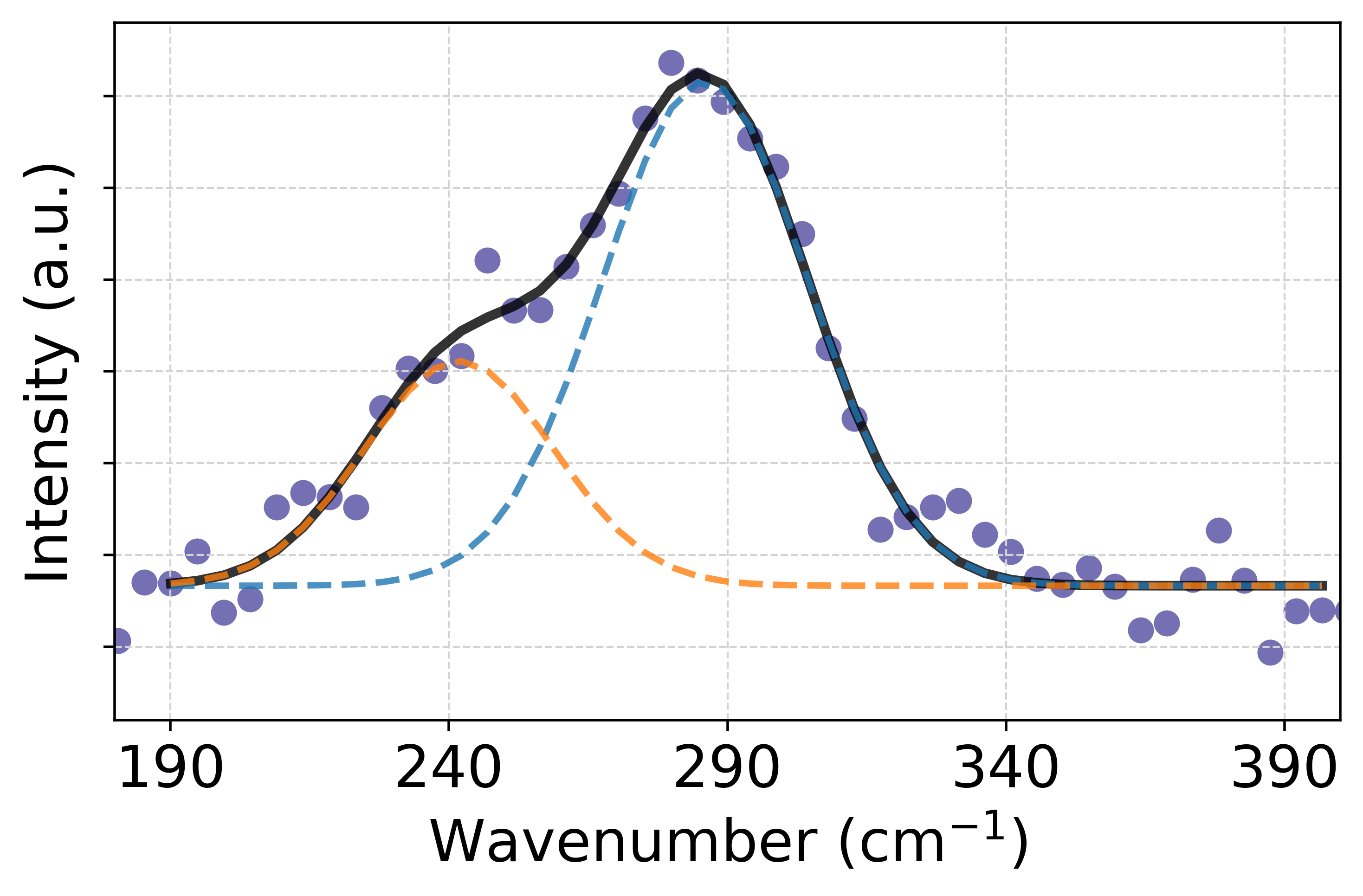}
    \caption{The low-frequency Raman peak of the 1ML superlattice is fitted by two Gaussian curves centered at 242 $\pm$ 17 cm\(^{-1}\) and 286 $\pm$ 18 cm\(^{-1}\) with a R$^2$-value of 0.0971.}
    \label{fig:ramanFit}
\end{figure*}

\begin{figure*}
    \centering
    \includegraphics[width=0.9
    \linewidth]{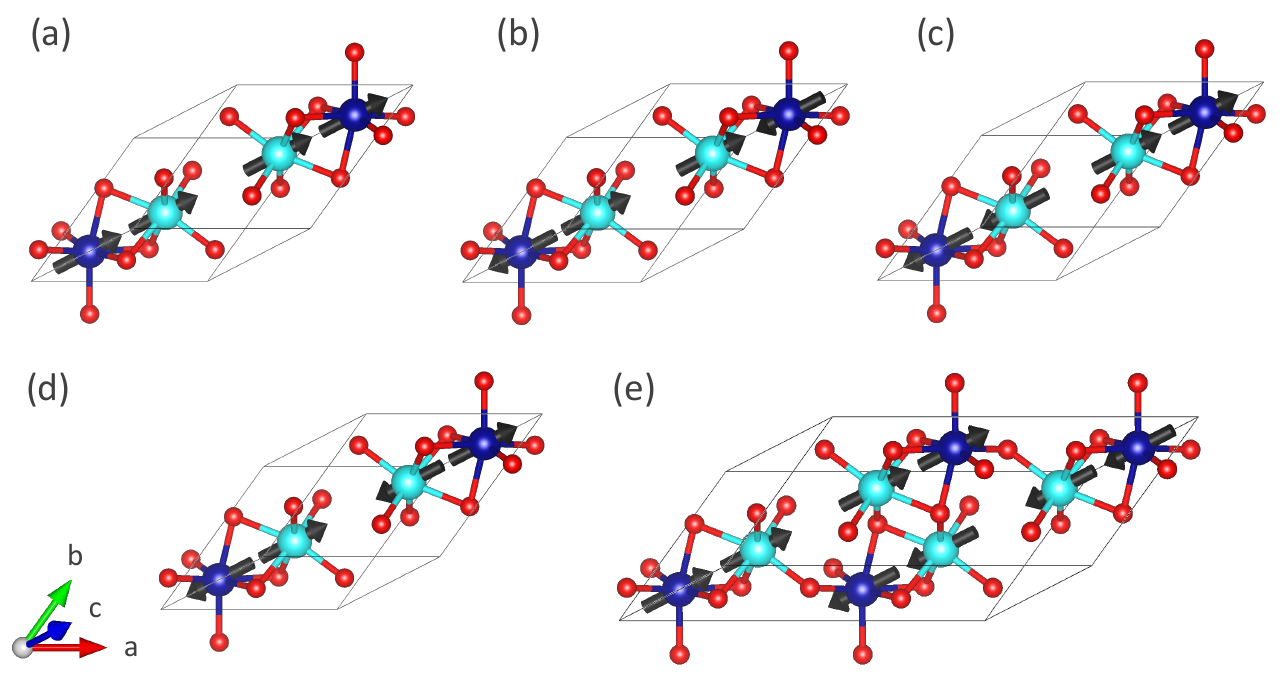}
    \caption{Magnetic ordering configurations in ilmenite CrVO$_3$. a) Ferromagnetic ordering, b) Ferrimagnetic ordering, c-e) Antiferromagnetic ordering.}
    \label{fig:magneticOrdering}
\end{figure*}

\begin{figure*}
    \centering
    \includegraphics[width=0.8\linewidth]{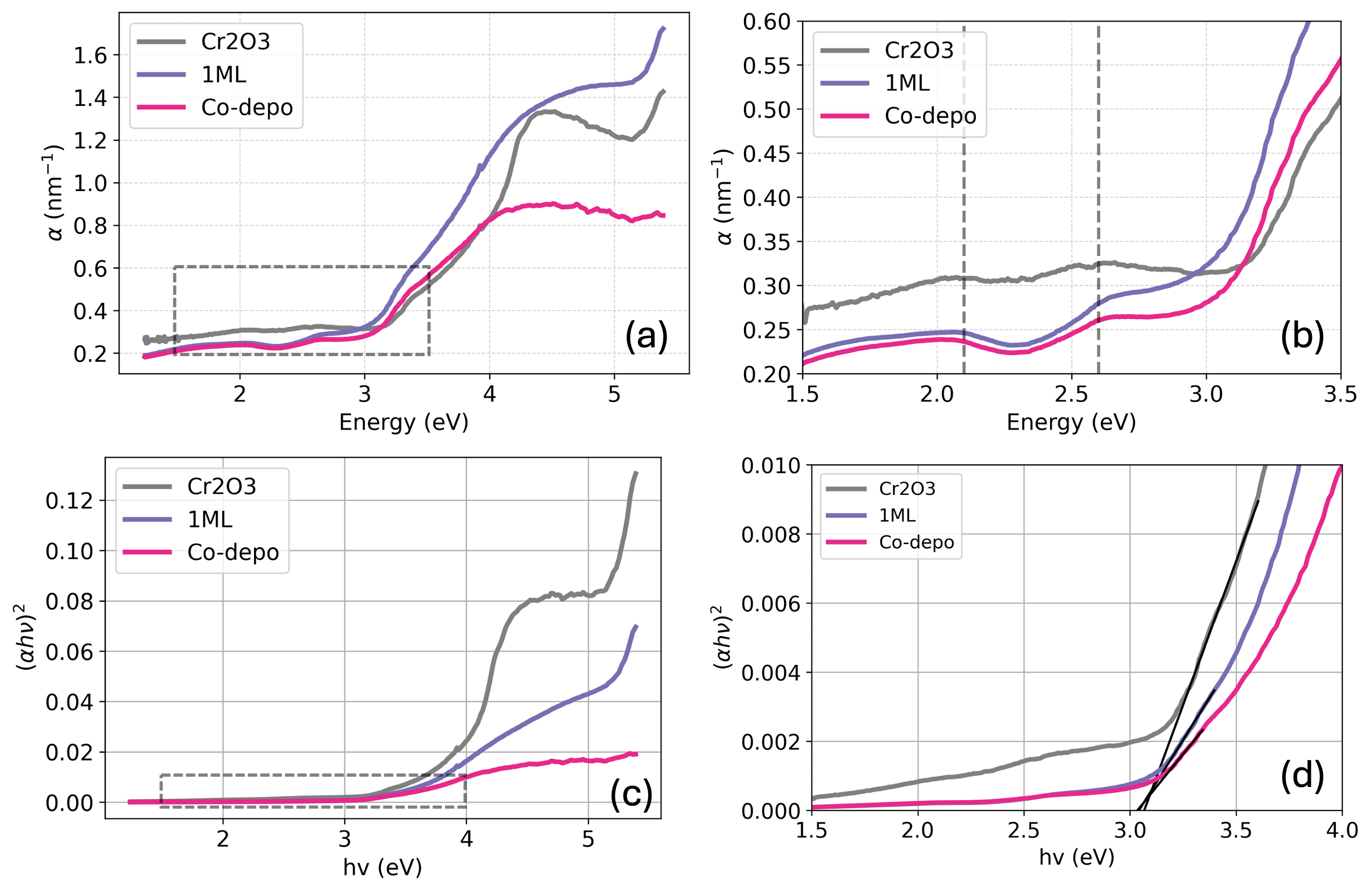}
    \caption{UV-VIS spectroscopy of the 1ML superlattice, co-deposited CrVO$_3$ and Cr$_2$O$_3$ thin films. a) Absorption coefficient as a function of photon energy across the full measured range. b) Magnified view of the region indicated by the dashed grey rectangle in (a). The two vertical dashed lines at 2.1 eV and 2.6 eV indicate the Cr$^{3+}$ 3d crystal field excitations. c) Tauc plot derived from the absorption coefficient data. d) Zoomed-in area from (c). The linear region of the curves is fitted and the x-intercept gives a similar band gap value for all the samples (\~3.1 eV). The fitted band gap value in the Tauc plot should be interpreted carefully, since the CrVO$_3$ samples are grown on top of Cr$_2$O$_3$ buffer layer. On the other hand, we note that the onset of the band gap absorption in (b) occurs at slightly lower energy for the CrVO$_3$ samples. This might suggest that both the 1ML superlattice and the co-deposited CrVO$_3$ have E$_g$<3.1 eV. However, a quantitative estimate of the band gap is hindered by the presence of the buffer layer band gap as well as the Cr$^{3+}$ absorption at 2.6 eV.}
    \label{fig:uvvis}
\end{figure*}

\begin{figure*}
    \centering
    \includegraphics[width=0.7\linewidth]{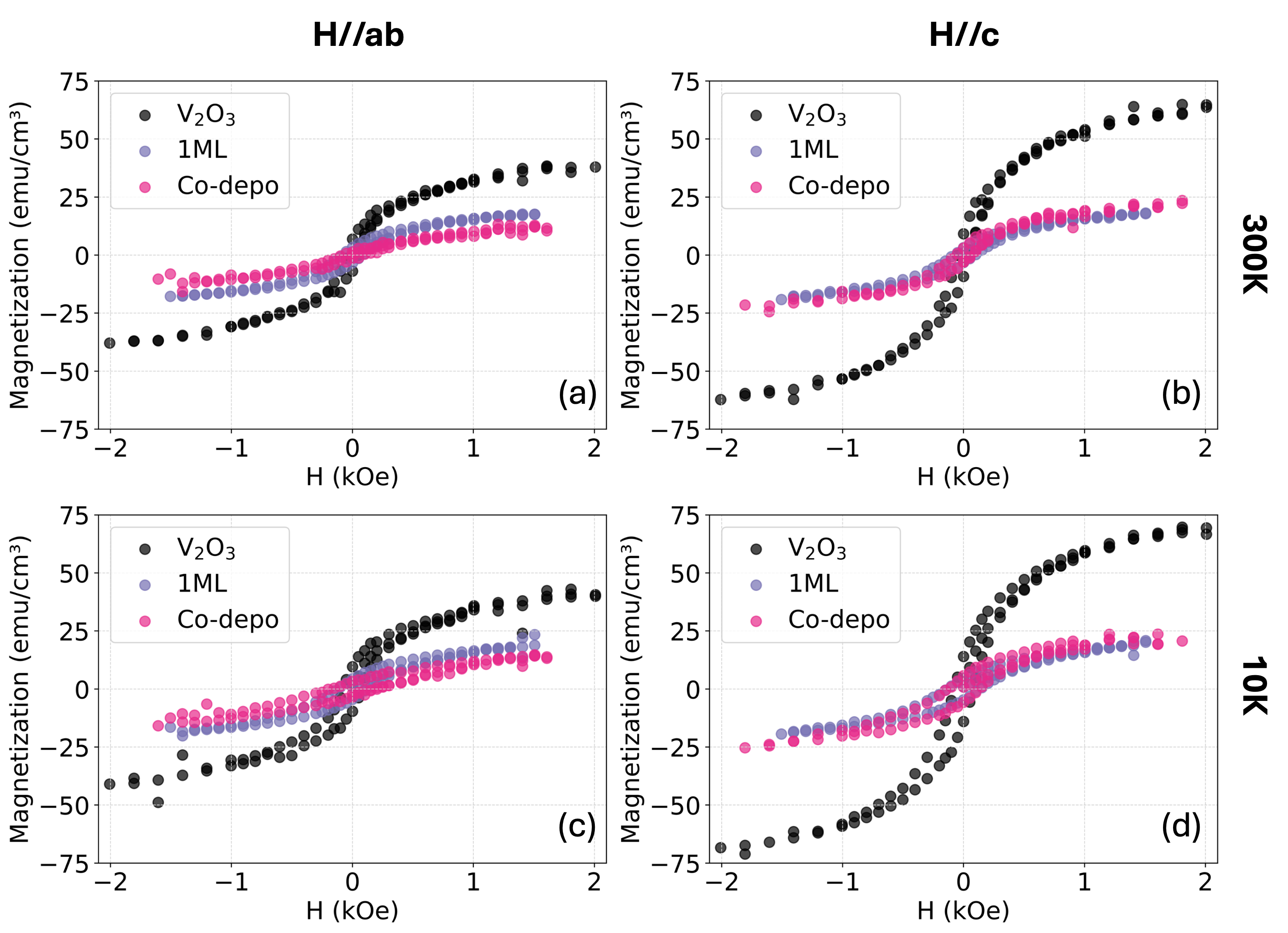}
    \caption{Magnetization versus the applied magnetic field H at 300 K (top row) and 10 K (bottom row). H is applied along the in-plane (H//ab, left column) or the out-of-plane direction (H//c, right column).}
    \label{fig:MvsH}
\end{figure*}

\end{document}